\documentclass[twocolumn,amsmath,showpacs,amssymb,aps,nofootinbib,floatfix]{revtex4-1}

\usepackage{epsfig,amsmath,amssymb}
\usepackage{dcolumn,color}
\usepackage{bm}
\usepackage{graphicx}
\usepackage{cleveref}
\usepackage{footnote}
\usepackage{mathtools, nccmath}
\usepackage{lmodern}
\begin{document}
\title{Generalized quantum electrodynamics: one-loop correction}
\author{David Montenegro$^1$}\thanks{david.montenegro@unesp.br}
\affiliation{$^1$S\~{a}o Paulo State University (UNESP), Institute for Theoretical Physics (IFT), R. Dr. Bento Teobaldo Ferraz 271 CEP 01140-070, S\~{a}o Paulo, SP.}
\begin{abstract}

In this paper, we give an update on divergent problems concerning the radiative corrections of quantum electrodynamics in $(3+1)$ dimensions. In doing so, we introduce a geometric adaptation for the covariant photon propagator by including a higher derivative field. This derivation, so-called generalized quantum electrodynamics, is motivated by the stability and unitarity features. This theory provides a natural and self-consistent extension of the quantum electrodynamics by enlarging the space parameter of spinor-gauge interactions. In particular, Haag's theorem undermines the perturbative characterization of the interaction picture due to its inconsistency on quantum field theory foundations. To circumvent this problem, we develop our perturbative approach in the Heisenberg picture and use it to investigate the behavior of the operator current at $1-$loop. We find the $2-$ and $3-$point correlation functions are ultraviolet finite, electron self-energy and vertex corrections, respectively. On the other hand, we also explain how the vacuum polarization remains ultraviolet divergent only at $e^2$ order. Finally, we evaluate the anomalous magnetic moment, which allows us to specify a lower bound value for the Podolsky parameter. 

\end{abstract}
\maketitle



\section{Introduction}

Fields are extremely powerful tools to establish a suitable language for various elementary physical phenomena that we have. As long as the physical phenomena are not known, fields themselves serve as a mathematical guide to elucidate our ignorance about the intrinsic features of nature laws. For instance, it occurs for the quantum electrodynamics $(QED_4)$, the most suitable theory to describe the photon-matter interaction. In any case, we can persuade nature to disclose a more fundamental description by looking for irregularity, analyzing experiments, and performing statistical methods on data. In this way, we have noted a significant conflict between the experimental and theoretical predictions from the standard model \cite{Ellis1998eh}. Among these deviations, we observe the muon anomalous magnetic moment \cite{deuxdeux} and Lamb shift in the hydrogen spectra \cite{unun}. Much of these processes encourage us to develop an alternative approach since the standard model is not mature enough to support a reasonable explanation. The discrepancies cited are opportunities to search a new physics beyond the standard model \cite{SMSM} if we believe the theoretical knowledge is imprecise. For realization of such task, there are important guidelines as the Higgs Model \cite{Deandrea}, SUSY \cite{SUSY}, Composite Higgs models \cite{composite}, and Effective field theory models \cite{comprendre}.

It is well-known that adding higher derivative corrections to the original theory is a procedure for constructing a more fundamental one. In recent years, effective field theory models have shown much progress in many physical contexts as modification of gravity \cite{clagrav}, cosmology \cite{cosmocosmo}, and string theory \cite{stringstring}. Moreover, including higher derivative terms in physical models goes beyond the pure mathematical exercises. The addition of these to standard equations of motion comes with benefits. The interesting consequence of such a study is that they can incorporate different energy scales. For example, these resulting systems can control the ultraviolet infinities \cite{1234} and provide singularity-free. In what concerns the further application, we note the three-dimensional gravity model arises a natural regulator, suppressing the ultraviolet divergence \cite{Deserasd}. In addition, we can resolve the singularity problems as the gravitational potential $r^{-1}$ at the origin in the Newtonian-limit \cite{Giacchini} and initial curvature in cosmological framework \cite{Easson}. In this work, we show how to draw a new perspective about underlying physical aspects of the $QED_4$. In general, one can explore the effective field theory method using a Lagrangian expansion in powers of derivatives, satisfying the standard model symmetries

\begin{equation}\label{socorro}
    \mathcal{L} =  \mathcal{L}_0  +  \sum_n \frac{C_{j,n}}{\Lambda^{n-4}} \mathcal{O}_{j,n},
\end{equation}
where the Wilson coefficients are $C_{j,n}$, a renormalizable and known Lagrangian is $\mathcal{L}_0$, and parameter $\Lambda$ encodes a new physical scale. The $\mathcal{L}_0$ and $\mathcal{O}_{j,n}$ are composed of standard model local fields.

Besides the quantum deviation already mentioned, we find evidence of the unsolved questions in electrodynamics at the classical level. Then if one adopts the $QED_4$ view as an inaccurate theory, we can create a favorable scenario to develop an alternative description. Podolsky developed, in 1942 \cite{Podolskyorigin}, the so-called generalized quantum electrodynamics $(GQED_4)$ by means of the philosophy in \eqref{socorro}. Podolsky proposed a second order gauge theory whose Lagrangian density is $ - \frac{1}{4} F^{\mu\nu}F_{\mu\nu} - \frac{1}{2m_P^2} \partial^\mu F_{\mu\beta} \partial_\alpha F^{\alpha\beta}$. By comparing with \eqref{socorro}, we have $\mathcal{L}_0$ as the $QED_4$, $C_{j,6}=1/2$, $\Lambda = m_P$, and $\mathcal{O}_{j,6}$ composed by $\partial^2 A$ terms. This theory fulfills the requirement of Lorentz and Gauge invariance and linearity (superposition principle). What motivates us to investigate is that the Podolsky version is unique up to a surface term among all other second order gauge theories at Lagrangian level \cite{Cuzinattogauge}. Furthermore, Podolsky solved, at classical level, the self-energy singularity $r^{-1}$ at the static charge \cite{podsel} and the famous $4/3$ problem \cite{43problem}. It is instructive to look into the question of whether or not the $GQED_4$ presents more unphysical degrees of freedom than the $QED_4$. Considering the vacuum state must be uniquely defined, we analyze what is the best-suited gauge fixing. At first glance, we may argue the Lorentz gauge as the best candidate $\Omega[A] = \partial_\mu A^\mu $ \cite{Podolskyorigin}. However, it is worth pointing out this divergenceless solution still identifies the photon with an equivalent class of longitudinal vector fields \cite{galvao}, whereas the transverse gauge components are fixed \eqref{era}. To restore the gauge invariance, we use the non-mixing gauge $\Omega[A] = \sqrt{1 - \frac{\Box}{m_P^2}} \partial_\mu A^\mu $ \cite{nomixing} whose the main characteristic is to preserve the uniqueness solutions of the equations of motion.

As we know, it is not surprising that the higher derivative theories present disastrous consequences for classical and quantum systems. This behavior yields, for example, problems associated with the lack of lower-energy bound, canonical quantization, negative energy states, violation of energy bounds on cross sections, and non-conservation of probability  \cite{ostroortso,Pais}. The main non-trivial step is to clarify what means stability concept. If we look more closely, the canonical Noether's energy is not a reliable approach to determine the stability \cite{russo}. Recently, to remedy this situation, Kaparulin \textit{et al} established a relevant connection between the time translational invariance and positive integral of motion by implementing the Lagrangian anchor concept. This study reveals the $GQED_4$ is stable \cite{russo}. In addition, the unitarity \cite{Pimente}, invariance under Becchi-Rouet-Stora-Tyutin $(BRST)$, proves the $GQED_4$ can no longer belong to the effective field theory class.

According to the conventional interpretation, the interaction picture is the basis for exploring non-trivial quantum phenomena. This standard perturbative formalism has archived great success in extracting physical predictions from cross section experiments. However, Haag's theorem proved several complications surrounding this picture, which manifest problems to the quantum field theory foundation \cite{Haag}. What ends up being the consequence is the non-Fock representation to describe interacting fields. In other words, the interaction picture cannot assume a Fock representation for asymptotic times, governed by the free Hamiltonian $H_0$. Furthermore, the interacting Hamiltonian $H_{int}$ cannot annihilate the free vacuum, where the elementary phenomena occur as vacuum polarization. As a consequence, there is no proportional relation between the dressed vacuum $ | \Omega \rangle \in \mathcal{H}_{int} $  and free one  $ | 0 \rangle \in \mathcal{H}_{0}$. Then we can assume the Hilbert space of interacting $\mathcal{H}_{int}$ and free $\mathcal{H}_{0}$ operators are disjoint. This statement has a consistent physical interpretation. We cannot relate a physical state $\in \mathcal{H}_{int}$ to a Fock space by a unitary equivalence transformation in the interaction picture.

The arguments above are a welcome development for quantum field theory in the Heisenberg picture, first used by Källén \cite{Kallen1,Kallen2}. This derivation asserts the roots for a perturbative formalism where the quantum field objects are valid in all Hilbert space. This method allows us to study quantum fluctuations with a Fock representation for interacting fields, where the interacting vacuum is invariant under Euclidean transformation. The traditional approach of the interaction picture is quite different from the Heisenberg picture. The former applies the Green's function as the main object to evaluate radiative corrections, whereas the latter expands the operators into a power series in gauge coupling, combined with the equations of motion. 




We understand the conservation of operator current $(j^\mu(x))$ arises from a continuous symmetry. One then considers this operator the main physical variable to study the $1-$loop behavior by writing its expansion in the Heisenberg picture. In this language, we show the quantum effects of the $2-$ and $3-$point correlation functions, the electron self-energy and vertex correction, respectively, are ultraviolet finite at all orders. In fact, the vacuum polarization contains ultraviolet divergence at $e^2$ order because its only virtual particles are fermions. Besides that, we can refine our physical interpretation by evaluating anomalous magnetic moment of electron $\frac{1}{2}(g-2) -\frac{\alpha}{2 \pi}$. In this case, we can settle down a new dispersion relation that establishes a lower bound on the Podolsky parameter $(m_P)$.

The layout of this paper is organized as follows. In section \ref{deux}, we introduce a brief study of the $GQED_4$ formalism. We show how Podolsky correction suggests two orthogonal families of the solution, which lead to finite results in the asymptotic state. This success is based on the higher derivative term, suppressing the ultraviolet divergence in the photon propagator. Section \ref{444}, somewhat technical, addresses the formalism of the radiative process under the Heisenberg picture formalism. We discuss our findings in section \ref{555}, more specifically, we calculate the $1-$loop electron self energy \ref{quatre} and vertex corrections \ref{trois} at $e^2$ order. We make contact with the anomalous magnetic moment in section \ref{fin}. Finally, section \ref{doendo} is left for the conclusion and outlook. More details are present in the Appendix \ref{appa}, \ref{appa2}, and \ref{apdc}.

\section{Formalism}\label{deux}

We introduce a quick review of the formal aspects of photon covariance in the $GQED_4$. Let us start this discussion with the Podolsky Lagrangian, defined by\footnote{We shall adopt the notation $ ds^2 = - dx_\sigma^2 $ where $x_4 = i x_0 = i c t$, $\hbar = c = 1$, and Greek indices run from $1$ to $4$.} 
 
\begin{equation}\begin{aligned}\label{llkk2}
\mathcal{L}_{GQED_4} &= -\frac{1}{4}F^{\mu\nu}F_{\mu\nu} - \frac{1}{2m_P^2} \partial^\mu F_{\mu\beta}\partial_\alpha F^{\alpha\beta} \\ 
& \qquad \qquad \qquad \qquad - \frac{1}{2 \xi}(1- \frac{\Box}{m_P^2})(\partial_\mu A^\mu)^2,
\end{aligned}\end{equation} 
where the strength field $F_{\mu \nu} = \partial_\mu A_\nu -\partial_\nu A_\mu$ satisfies the Bianchi identity and we use the Feynman gauge $\xi = 1$. The covariant Euler-Lagrange equations become

\begin{equation}\label{elpod}
(1- \frac{\Box}{m_P^2} ) \Box A^{\mu} (x) = 0.
\end{equation}
Our higher derivative theory increases the families of the solution, so these new degrees of freedom create distinctive features which will be helpful to control the ultraviolet behavior. We can recover the free $QED_4$ solutions from the free Podolsky ones. This property, easily seen by taking $m_P \rightarrow \infty$, preserves the Lorentz and gauge invariance and linearity features.

It may be helpful to write the gauge field as the product of two orthogonal ones. Thus, we divide the degrees of freedom by applying $A^\mu (x) = A_{Max}^\mu (x) + A_{Pod}^\mu (x)$ in the equations above 

\begin{equation}\begin{aligned}\label{era}
(1 - \frac{\Box}{m_P^2} ) A^\mu (x) &= A_{Max}^\mu (x), \ \    \frac{\Box}{m_P^2} A^\mu (x) =  A_{Pod}^\mu (x),
\end{aligned}\end{equation}
where $A_{Max}^\mu (x)$ and $A_{Pod}^\mu (x)$ are the Maxwell and Podolsky gauges, respectively. One reads 

\vspace{1mm}

\begin{equation}\label{cachorro}
\Box A_{Max}^\mu (x) = 0, \quad (\Box - m_P^2 ) A_{Pod}^\mu (x) = 0,
\end{equation}
where $A^\mu_{Max} (x)$ and $A^\mu_{Pod} (x)$ obey the Maxwell electrodynamics and "Proca" differential equations, respectively. This separation, only valid in the free case, explains why we may interpret $A_{Pod}^\mu (x)$ as a massive gauge\footnote{In the section \ref{555}, we proved $m_P$ has a cutoff interpretation to deal with divergences.}. The gauge invariant relations at different time are

\vspace{1mm}

\begin{equation}\begin{aligned}\label{qwe2}
[A_\mu (x), A_\nu (x')] &= -i \delta_{\mu\nu} G (x'-x), \\ \ 
\langle 0 | \{ A_\mu (x), A_\nu (x')  \} | 0 \rangle &=  \delta_{\mu\nu} G^{(1)} (x'-x), 
\end{aligned}\end{equation}
where a careful inspection shows 

\vspace{1mm}

\begin{subequations}\begin{align}\label{sssa}
G (x) &= - i \int \frac{d^4 p}{(2\pi)^3}  e^{ipx} \epsilon(p) ( \delta^4 (p^2 ) - \delta^4 (p^2 + m_P^2)),\\ \label{sssa2}
G^{(1)} (x) &= \int \frac{d^4 p}{(2\pi)^3}  e^{ipx} ( \delta^4 (p^2) - \delta^4 (p^2 + m_P^2)).
\end{align}\end{subequations}
For latter application in perturbation theory, we split, by definition, the distributions \eqref{sssa} into two different regions: the backward $G_R (x) = - \Theta(x_0) G(x)$ and forward $G_A (x) = \Theta(- x_0) G (x)$ light cone\footnote{We define the Heaviside function as $ \Theta(p) \equiv (1 + \epsilon(p))/2 $ with the sign function $\epsilon(p) \equiv p_0 /|p|$.}. The singular functions above are composed of two orthogonal hyperboloid surfaces. This geometric adaptation allows to cancel the divergences in the interacting system \cite{insuportavel}.

\section{Heisenberg perturbative method}\label{444}

In this section, we develop the perturbative approach proposed by Kállën \cite{Kallen1,Kallen2} to evaluate the quantum fluctuations. The analytical propagators derived in the previous section will be sufficient for our purposes. Starting with the Lagrangian of \eqref{llkk2}, free Dirac, and minimal coupling, we have

\begin{equation}\begin{aligned}\label{asdfgfds}
\mathcal{L} &=-\frac{1}{4}F^{\mu\nu}F_{\mu\nu} - \frac{1}{2 m_P^2} \partial^\mu F_{\mu\beta} \partial_\alpha F^{\alpha\beta} + j_\mu A^\mu \\
& \quad - \frac{1}{2}(1-\frac{\Box}{m_P^2})(\partial_\mu A^\mu)^2 - \frac{1}{4} [\bar{\psi}, \ ( \gamma \cdot \partial + m ) \psi ] \\
& \quad - \frac{1}{4} [\bar{\psi} ( \gamma \cdot \overleftarrow{\partial} + m ), \ \psi ].
\end{aligned}\end{equation}
We note the Podolsky's theory preserves the minimal coupling and free Dirac terms of the $QED_4$ Lagrangian to maintain the Lorentz invariance \cite{Podolskyorigin}. We redefine the current operator as

\begin{equation}\label{j0}
j_{\mu} (x) \equiv \frac{ie}{2} [\Bar{\psi}(x), \gamma_\mu \psi(x)],
\end{equation}
which is related with $ j_{\mu} (x) = \ : \overline{\psi} (x) \gamma_{\mu} \psi (x): $ by Wick's theorem \cite{Bogoliubov}, and consequently $\langle 0|j_\mu(x)|0 \rangle = 0$. Even though the minimal interaction has dimension of coupling equal to zero, the $GQED_4$ is a super-renormalizable theory\footnote{This interesting difference is analyzed in the Appendix \eqref{apdc}.}. The dynamical equations from variational principle read

\begin{equation}\label{cde2}
(\gamma \cdot \partial + m)  \psi(x) = i e \gamma^\mu A_\mu (x) \psi (x), 
\end{equation}
\begin{equation}\label{cde3}
(1 - \frac{\Box}{m_P^2}) \Box A_\mu (x) = -\frac{ie}{2} [\Bar{\psi} (x) \gamma_\mu , \psi (x)].
\end{equation}
As we mentioned at the beginning of this section, we shall give a different meaning to the splitting method of the interaction picture. To do so, we can rewrite in a slightly different manner \cref{cde2,cde3} as the integral equations of motion, preserving all symmetries

\begin{subequations}\begin{align}\label{p1}
\psi(x) &= \psi^{(0)}(x) - \int d^4 x'  S_R (x-x')  i e \gamma^\mu A_\mu (x') \psi(x'), \\
\label{p3}
A_\mu (x) &= A^{(0)}_\mu (x) - \int d^4 x' G_R (x-x')  \frac{ie}{2} [\Bar{\psi} (x') ,\gamma_\mu \psi(x')],
\end{align}\end{subequations}
where $S_R (x-x')$ is the retarded fermion propagator, defined in \cite{schwinger} and $(\psi^{(0)} (x) , A^{(0)}_\mu (x) )$ are the free operators. Quite naturally, it is possible to consider the advanced singular function to write down these solutions. The interacting operators $(\psi (x), A_\mu (x))$ are the sum of homogeneous and inhomogeneous solutions of the differential \cref{cde2,cde3}. We argue, contrary to the interaction picture, the sum of each operator is well-defined separately by canonical commutation relations. Indeed, the transformation under the Euclidean group $E(4)$ holds at all times because the free and interacting operators live in the same Hilbert space, and thus the unitary transformation is guaranteed. We can therefore construct an irreducible Fock representation, relating the creation and annihilation operators at asymptotic time $( x_0 \rightarrow \pm \infty )$ with the finite one. Moreover, we stress that the incoming and outgoing operators, responsible for the physical states, are deeply connected with $(\psi^{(0)} (x) , A^{(0)}_\mu (x) )$ since they satisfy the same commutation relations \cite{Yang}. The Heisenberg picture takes the Hilbert space from total Hamiltonian $ H = H_0 + H_{int}$ as the basis where the interacting fields can provide an irreducible representation of canonical commutation relations at the equal time. All these assumptions assure a proper definition for a Fock space.

To construct the perturbation-theory apparatus, we express the operators into a series expansion in the power of gauge coupling where each subsequent term is sub-leading to the previous one. We justify the electric charge as the possible coupling because of smallness $e^2/4\pi \sim 1/137$,

\begin{equation}\begin{aligned}\label{t1}
\psi(x) &= \psi^{(0)}(x) \ + \ e\psi^{(1)}(x) \ + \ e^{2}\psi^{(2)}(x) \ + \ldots, \\ 
A_{\mu}(x) &= A_{\mu }^{(0)}(x) \ + \ e A_{\mu }^{(1)}(x) \ + \ e^{2}A_{\mu}^{(2) } (x) \ +\ldots. \\ \\ \\ 
\end{aligned}\end{equation}
Our principal aim is to express the higher order operators explicitly as a function of the free field operators. Considering terms with the same power of $e$, we get the general recursive relations for any arbitrary approximation. If we replace \eqref{t1} into \cref{cde2,cde3}, we obtain

\begin{subequations}\begin{align}\label{RRpsi}
\psi^{(n+1)}(x) &=-\frac{i}{2}\int{d^4x'}  S_R(x-x') \gamma^\mu \times \nonumber \\
&\quad{} \qquad  \quad \sum_{m=0}^{n} \{ A_{\mu }^{(m)}(x'),\psi^{(n-m)}(x') \}, \\
\label{RRApsi}
A_{\mu }^{(n+1)}(x) &=\frac{i}{2}\int{}d^{4}x'  G_R (x-x') \times \nonumber \\
& \qquad \quad \quad  \sum_{m=0}^{n} [\bar{\psi}^{(m)}(x')\gamma_{\mu},\psi^{(n-m)}(x') ].
\end{align}\end{subequations}
We require a similar argument used in the discussion above. One expands the observable quantities in a quite analogous way with \eqref{t1}. Following this procedure, we focus on the current operator expansion in a series power of $e$ to determine the quantum available physical process

\vspace{3mm}

\begin{equation}\label{EXPCurren}
j_{\mu} (x) = j^{(0)}_{\mu} (x) + e j^{(1)}_{\mu} (x) +  e^2 j^{(2)}_{\mu} (x) + \ldots.
\end{equation}
To address only the transition amplitude with one internal loop, we shall evaluate the current operator at $e^2$ order by substituting \eqref{t1} into \eqref{j0} with the aid of \eqref{RRpsi} and \eqref{RRApsi}


\begin{widetext}

\hfill

\begin{equation}\begin{aligned}\label{0j2}
j^{(2)}_{\mu} (x)  
&= \ \frac{i}{8}\int{d^4x'}\int{d^4x''} [ \bar{\psi}^{(0)}(x),\gamma_{\mu} S_R(x-x')\gamma_{\nu_1} \{ \psi^{(0)}(x') , [ \bar{\psi}^{(0)}(x''), \gamma^{\nu_1} \psi^{(0)}(x'') ] \} ]  G_R(x'-x'')\\
& \  - \frac{i}{4}\int{d^4x'}\int d^4x'' [ \bar{\psi}^{(0)}(x), \gamma_{\mu} S_R(x-x') \gamma^{\sigma_1} S_R(x'-x'')  \gamma^{\sigma_2} \psi^{(0)}(x'') ] \{ A_{\sigma_1}^{(0)}(x') , A_{\sigma_2}^{(0)}(x'')   \} \\
& \ - \frac{i}{4} \int{d^4x'} \int d^4x'' [ \bar{\psi}^{(0)}(x') \gamma^{\rho} A^{(0)}_{\rho} (x') S_A (x'-x), \gamma_\mu S_R (x - x'') \gamma^\beta A_\beta^{(0)} (x'') \psi^{(0)}(x'') ]  \\
&  \  + \frac{i}{8}\int{d^4x'} \int{d^4x''}[ \{ [\bar{\psi}^{(0)}(x''), \gamma_{\nu_2} \psi^{(0)}(x'') ], \psi^{(0)}(x') \} \gamma^{\nu_2}  S_A(x'-x) \gamma_\mu, \psi^{(0)}(x)] G_R(x'-x'') \\
& \ -\frac{i}{4}\int{d^4x'}\int{d^4x''} [ \bar{\psi}^{(0)}(x'')\gamma^{\alpha_2}S_A(x''-x')\gamma^{\alpha_1}S_A(x'-x),\gamma_{\mu}\psi^{(0)}(x) ] \{ A_{\alpha_1}^{(0)}(x') , A_{\alpha_2 }^{(0)}(x'') \}.
\end{aligned}\end{equation}
We shall only consider the situations where the transition amplitude cannot be precisely carried out by the conservation overall external energy-momentum. Then the matrix elements which fulfill this condition are $\langle q |j^{(2)}_{\mu} (x) | q' \rangle$ and  $ \langle 0 |j^{(2)}_{\mu} (x) | q, q' \rangle $. Thus

\begin{equation}\begin{aligned}\label{0nnn1}
e^3 \langle q |j^{(2)}_{\mu} (x) | q' \rangle
&= \frac{ie}{2}   \int {d^4x'} d^4x'' \langle q | [ \bar{\psi}^{(0)}(x) , \gamma_{\mu} S_R(x-x') \sum(x' - x'') \psi^{(0)}(x'')  ] |q' \rangle \ + \\
&\quad \ \frac{ie}{2}  \int {d^4x'} d^4x'' \langle q | [ \bar{\psi}^{(0)}(x'') \sum(x'' - x')   S_{A}(x'-x), \gamma_\mu \psi^{(0)}(x) ] | q' \rangle \ + \\
&\quad \ \frac{ie}{2} \int{d^4x'} d^4 x''  \langle q | [ \bar{\psi}^{(0)}(x'), \Gamma_\mu (x'-x, x - x'') \psi^{(0)}(x'')] | q' \rangle \ + \\
&\quad \ \frac{ie}{2} \int{d^4x'} d^4 x'' \Pi_{\mu\nu} (x - x')  G_R(x'-x'') \langle q |  [\bar{\psi}^{(0)}(x''), \gamma^{\nu} \psi^{(0)}(x'') ] | q' \rangle. \\
\end{aligned}\end{equation}
\end{widetext}
We can separate the current operator into three particular contributions corresponding to well known radiative process in the $QED_4$. Strictly speaking, this matrix element represents the sum of electron self-energy $\sum $, vertex correction $\Gamma_\mu$, and vacuum polarization $\Pi_{\mu\nu}$.


\section{Radiative corrections in second order}\label{555}

During the development of the $QED_4$, the appearance of divergent expressions became an issue on its consistency. It was sufficient to yield unclear predictions for the vacuum expectation value of operators. We only have a well-established $QED_4$ when the regularization procedure gave a satisfactory method to deal with ultraviolet divergences.

In the preceding section, we developed a program to describe the spinor electrodynamics, where we can overcome the higher energy problems through a gauge invariant scheme of subtracting \eqref{sssa} and \eqref{sssa2}. Although we can turn the radiative corrections into finite, we must use the renormalization techniques to replace the bare coupling with the renormalized ones. The theoretical justification for such a method is the stability and unitarity of $GQED_4$, which encodes all symmetries from the $QED_4$. We propose to look at the higher derivative term, encoded by a negative mass dimension, in \eqref{asdfgfds} with caution. We cannot interpret this term as a non-renormalizable interaction but rather a kinetic contribution. In each order of perturbation theory, the new structures of $GQED_4$ are the same as the original one and have a closed number. Otherwise, if $GQED_4$ is non-renormalizable, news structures would arise when we add more loops. The Podolsky contribution resumes affecting the kinematics of the photon propagator. This term improves the convergence of the radiative process in the ultraviolet regime and keeps the number of counterterms finite as well as $QED_4$.

Following the ideas addressed so far, the vacuum polarization, one of the most elementary virtual phenomena, preserves at $1-$loop the same result of the $QED_4$\cite{QEDTomazelli}\footnote{ To illustrate this statement, we analyze the superficial degree of divergence in the Appendix \eqref{apdc}.}. It is crucial to point out the higher derivative term turns $GQED_4$ into a super-renormalizable theory, namely, there is a limited number of divergent amplitudes. Within the perturbative analysis, this observation yields one-particle irreducible of vacuum polarization is logarithmic divergent at $1-$loop, whereas it is ultraviolet finite at second order loop or more. In the next subsections, we focus on the analytical behavior of $\Gamma_\mu$ and $\sum$ functions at $e^2$ order and how the Podolsky covariant photon propagator removes these ultraviolet divergences, as in Table \ref{table1}. The Heisenberg picture brings new insights to interpret these radiative corrections.


\subsection{Electron self-energy $\sum$ }\label{quatre}

Our perturbative strategy relies on the Dirac equations of motion instead of Green's functions to investigate the electron self-energy. This method takes into account the quantum fluctuation by a subtle redefinition of the parameters from the equations of motion. We also develop the spectral representation, which retains the $1-$loop correction for the spin state. As a preliminary, we examine the matrix element expansion in \eqref{t1}

\begin{equation}\begin{aligned}\label{firstese}
&\langle 0 |\psi(x) | q \rangle \\
&= \langle 0 |\psi^{(0)}(x) | q \rangle +  e  \langle 0 |\psi^{(1)}(x)| q \rangle +  e^{2}  \langle 0 |\psi^{(2)}(x)| q \rangle +\ldots, \\ 
&= \langle 0 |\psi^{(0)}(x) | q \rangle + \int d^4 x' S_R(x-x')\langle 0 |\Phi(x') | q \rangle + \ldots,
\end{aligned}\end{equation}
where the first order $\langle 0 | \psi^{(1)}(x) | q \rangle$ naturally drops out. Here, we consider the lowest order of $e^2$ and multiplying \eqref{firstese} on both sides by the Dirac equation 

\begin{equation}\label{casado}
(\gamma \cdot \partial + m )  \langle 0 |\psi(x) | q \rangle = -  \langle 0 | \Phi (x) | q \rangle.
\end{equation}
Let us consider the qualitative behavior of the operator $\Phi (x)$\footnote{We place the explicit calculation involving $\Phi(x)$ in appendix \eqref{appa}.} under the perspective of the adiabatic hypothesis for which we introduce through the parameter $g$

\begin{equation}\begin{aligned}\label{hhh}
&\langle 0 |\Phi (x,g) | q \rangle = i \int \frac{d^4p}{(2\pi)^3}   \int^x_{-\infty} d^4 x'  e^{g(x_0 + x_0') + ip(x-x')}  \\ 
& \quad \ \times \epsilon(p) \bigg[ \sum_1(p^2) + (i\gamma \cdot p + m) \sum_2(p^2) \bigg]  e^{i q x'} u(q), 
\end{aligned}\end{equation}
where the factor $e^{g(x_0 + x_0')}$ allows the self-interaction to be adiabatically switched off. Now we get the spectral energy distribution

\begin{equation}\begin{aligned}
& \langle 0 |\Phi (x,g) | q \rangle = i e^{2 g x_0 + iqx} \int \frac{dp_0}{g + i(p_0 - q_0)}  \epsilon(p) u(q) \\
& \times \bigg[ \sum_1(\vec{q}^2 - p^2_0) +  (i\gamma^k p_k - \gamma^4 p_0  + m) \sum_2(\vec{q}^2 - p^2_0) \bigg] ,
\end{aligned}\end{equation}
with change of variable $p^2_0 = s + \vec{q}^2$, we obtain 

\begin{equation}\begin{aligned}\label{435534}
& \langle 0 |\Phi (x,g) | q \rangle = \ e^{2 g x_0 } \int^{\infty}_0   \frac{ ds}{s + \vec{q}^2 -(q_0 + ig)^2}  \\
&   \quad \quad \times \bigg[ \sum_1(-s) - i g \gamma_4 \sum_2(-s) \bigg] \langle 0 | \psi^{(0)}(x) | q \rangle.
\end{aligned}\end{equation}
To have a defined asymptotic state in the Fock space, we must perform the so-called adiabatic limit $g \rightarrow 0$. As consequence, the function argument $s$ goes to $m^2$. This operation arrives at a contradiction in the interaction picture if we take this adiabatic limit. As the dressed vacuum $| \Omega \rangle $ is proportional to the bare one $| 0 \rangle$, the total Hamiltonian cannot annihilate the Fock vacuum $H | 0 \rangle = 0 $, where vacuum polarization occurs. Therefore, this operation is inconsistent because of the self-interaction processes. We arrive at

\begin{eqnarray}\label{cacilda} 
\langle 0 |\Phi (x) | q \rangle =  \bar{ \sum_1} (-m^2) \langle 0 |\psi^{(0)}(x)| q \rangle,
\end{eqnarray}
where $\bar{ \sum_1} $ is the real part of $\sum_1$\footnote{See Eq. \eqref{lajel} for a better explanation.}. We shall clarify that the adiabatic switching off provides a smooth connection between the interacting operator and the asymptotic one. Such innocent operation may appear irrelevant. However, it captures the physical principle to obtain a meaningful vacuum state, which coincides with the Fock non-particle state \cite{Haagbook}. This limit only takes place in the Heisenberg picture where the sum of the free and interacting operator or, more precisely, the sum of respective solutions \cref{p1,p3} are well-defined in the same Hilbert space. On the other hand, the asymptotic and interacting operators from the interaction picture lie in inequivalent Hilbert spaces $\mathcal{H}_{in,out} \cap \mathcal{H}_{int} = \O $. Under this circumstance, we can straightforward obtain the same result above if we compute  \eqref{casado} on the mass shell value $p^2 = -m^2$

\begin{equation}\begin{aligned}\label{cdzqed}
& (\gamma \cdot \partial + m )  \langle 0 |\psi (x) | q \rangle \\ 
&= - \bigg[  \bar{\sum_{1}} (-m^2) + i \epsilon(q) \sum_1(-m^2) +(i\gamma \cdot q + m)  \times\\
& \qquad  \bigg(   \bar{\sum_{2}}(-m^2)  + i  \epsilon(q) \sum_2(-m^2) \bigg) \bigg] \langle 0 |\psi^{(0)}(x) | q \rangle,  \\ 
& = -  \bar{\sum_{1}}(-m^2)  \langle 0 |\psi^{(0)}(x)| q \rangle,
\end{aligned}\end{equation}
the difference between \eqref{cacilda} and \eqref{cdzqed} enable us to establish proof of the adiabatic limit validity for a massive theory with mass $ > 0$. The agreement with \eqref{cacilda} (asymptotic solution) is very impressive and shows the same spectral support corresponding to \eqref{cdzqed} (exact result). We now have this spinor equation that describes a single fermion where the parameters $\bar{\sum}_{1} $ and $\langle 0 |\psi (x) | q \rangle$ take into account quantum fluctuations effects at $e^2$ order. We may rewrite the foregoing equation more convenient as following

\begin{equation}\label{tre}
(\gamma \cdot \partial + m  +  \bar{\sum_{1}} (-m^2) ) \langle 0 |\psi (x) | q \rangle = 0,
\end{equation}
where the bare mass $(m)$ is a parameter in the free Dirac equation at $x_0 \rightarrow - \infty$. Using on shell subtraction scheme, we assume the $ \bar{\sum}_{1}$ corresponds to the pole mass $(m_{pole})$ minus $m$. However, the interaction picture undermines such an operation since this picture is responsible for a local rather than a global unitarity transformation. The immediate consequence is the absence of a unitary map  $S: \mathcal{H}_{in} \rightarrow \mathcal{H}_{out}$ for which we cannot formulate the asymptotic Hamiltonian $ H (A^{out}, \psi^{out}) = S^{-1} H (A^{in}, \psi^{in}) S $, where $S$ is the $S-$matrix \cite{Yang}. Therefore, we cannot demand a unitary operator from the Euclidean group, which transforms covariantly the states $m$ and $m_{pole}$ because both of them lie in the inequivalent Fock representation \cite{chatooo,cacilda2}. Formally, such an operation shows the interaction picture is completely inconsistent with a proper interpretation of quantum field theory \cite{Haagbook,Bogoljubov2}.

The adiabatic hypothesis plays a fundamental role, as of the utmost importance, to separate the interacting and asymptotic states because both of them are composed of the free operator. Returning to \eqref{tre}, we have

\begin{equation}\label{tre3}
(\gamma \cdot \partial + m_{pole})\langle 0 |\psi^{(r)} (x)|q \rangle=0,
\end{equation}
where $\psi^{(r)} (x)$ is the renormalized operator\footnote{Here, a spin state with a definite momentum $\gamma \cdot \partial$ and mass $m_{pole}$. Thus, the equation \eqref{tre3} leads to a propagator having a single pole at $\gamma \cdot p = - m_{pole}$ with unit residue.}. Then we completely characterize the electron self-energy process. A quite different way to understand \eqref{tre3} is by adding the counterterms $\delta_m$ in both sides of \eqref{cde2}

\begin{equation}\label{0zzzz}
(\gamma \cdot \partial + \underbrace{ m + \delta_m }_{m_{pole}} ) \psi(x) = i e \gamma \cdot A(x) \psi(x) + \delta_m  \psi(x), 
\end{equation}
one may be tempted to consider that this equation remains unchanged from the mathematical viewpoint. The $ \delta_m$ draws distinct physical concepts. It characterizes the "radiative mass", the bare pole $(m)$ is shifted by the quantum fluctuation to the pole mass $(m_{pole})$, if $e \rightarrow 0$, $\delta_m \rightarrow 0$ in the right side. This process is called \textit{mass renormalization} in the left side. In fact, we obtain \eqref{cdzqed} at $e^2$ order if we require the mass counterterm as

\begin{equation}\label{ffpp}
\delta_m = -  \bar{\sum_{1}}(-m^2),
\end{equation}
it follows from \eqref{0htrans11} that  

\begin{equation}\begin{aligned}\label{countmass}
\delta_m = \frac{ e^2}{16 \pi^2} &  \bigg[ 1  +  \frac{12}{b}\bigg( \frac{b + 1 }{ b - 1} \bigg)^2 \ln{\bigg(\frac{2b}{b^2 + 1}\bigg)} -   \\
&   \bigg( 2 + \frac{1}{b^2}  \bigg)  \sqrt{\frac{ 1 + 2 b}{ 1 - 2 b }}  \ln{\bigg( \frac{1- \sqrt{\frac{ 1 - 2 b}{ 1 + 2 b}}}{1+ \sqrt{\frac{ 1 - 2 b}{ 1 + 2b }}} \bigg)} \bigg],
\end{aligned}\end{equation}
where $b= \frac{m}{m_P}$. The subtraction procedure between the non-massive and massive spectrum cancels the infinities from the $1-$loop integrals\footnote{Podolsky produces two orthogonal solutions $\delta^4 (p^2)$ and $\delta^4 (p^2 + m^2)$ on-shell conditions. The off-shell intermediate states combine these hypersurfaces in momentum space integral. We note \eqref{demandé} to illustrate this point.}. In short, if we take $m_P \rightarrow \infty$, the results of $GQED_4$ does not converge to $QED_4$. Therefore, this notable characteristic shows we cannot take $m_P$ as a small correction but rather responsible for new solutions, improving the behavior in the ultraviolet regime. Finally, we will discuss the $\psi^{(r)} (x)$. A further examination in \eqref{tre3} shows the matrix element is

\begin{equation}\label{5577}
\langle 0|\psi^{(r)} (x)|q \rangle = Z_2^{-1/2} \langle 0|\psi^{(0)} (x)|q\rangle,
\end{equation}
where $Z_2^{-1/2}$ is an infinity number requested to normalize the wave function. This relation corresponds to the physical ("dressed") electron state in the function of the bare one. To obtain the correct $Z_2$, we cannot solve the operator \eqref{hhh} directly without subtracting by the asymptotic state \eqref{cacilda}\footnote{ The $\Phi (x,g)$ leads to the asymptotic state \eqref{cacilda} in the limit $g \rightarrow 0$.}

\begin{eqnarray}\begin{aligned}\label{label}
& \langle 0 |\Phi (x,g) | q \rangle -  e^{2 g x_0 } \bar{ \sum}_1(-m^2) \langle 0 |\psi^{(0)}(x)| q \rangle =  \\
& e^{2 g x_0 } \int ds \bigg[ \frac{ (2i q_0 g -g^2) \sum_1(-s) }{(s-m^2)(s + \vec{q}^2 -(q_0 +ig)^2)} \\
& \qquad \qquad \ \ - \frac{i \gamma_4 g \sum_{2}(-s)}{s+\vec{q}^2 -(q_0 +ig )^2 } \bigg] \langle 0 |\psi^{(0)} (x)| q \rangle.
\end{aligned}\end{eqnarray}
To make a proper subtraction, we insert the factor $e^{2 g x_0}$ due to the fact $\bar{ \sum}_1 (-m^2)$ depends on the gauge coupling. Moreover, $\bar{ \sum}_1 (-m^2)$ plays the role of the "radiative mass" (self-mass), the additive contribution due to photons interacting with the bare electron. The physical meaning of this subtraction is to obtain the physical ("dressed") electron state. At this stage, we evaluate the correct effect for the renormalized propagator by removing the change in the electron mass. The consistency for such an operation is the adiabatic limit which smears from the interacting operator to the asymptotic one by a test function. This physical interpretation, founded on the adiabatic hypothesis, concedes that we can define a set of well-adjoint operators on the Fock space. Substituting \eqref{label} into \eqref{firstese}, the matrix element expansion becomes

\begin{equation}\begin{aligned}
&\langle 0 |\psi^{(r)} (x) | q \rangle = \\ 
&  \langle 0 |\psi^{(0)}(x)| q \rangle + i \int \frac{d^4 p}{(2 \pi)^3} \int^{x}_{-\infty} d^4 x' e^{ip(x-x')} (i \gamma p - m)  \\
&  \times \epsilon(p) \delta^4 (p^2+ m^2)  e^{2 g x'_0 }   \int^{\infty}_0 ds \bigg[ \frac{(2i q_0 g -g^2)}{(s + \vec{q}^2 -(q_0 +ig)^2)}   \\
&  \times \frac{\sum_1(-s)}{(s-m^2)} - \frac{ i \gamma_4 g \sum_2 (-s)}{s+\vec{q}^2 -(q_0 +ig )^2 } \bigg] u(q) e^{i q  x'}.
\end{aligned}\end{equation}
After integrating over the space and momentum, we have

\begin{equation}\begin{aligned}
& \langle 0 |\psi^{(r)} (x)  | q \rangle = \\
& \langle 0 |\psi^{(0)}(x)  | q \rangle \bigg\{ 1  +  \frac{e^{2g x_0}}{4 q_0 } \bigg[ \frac{(i \gamma \cdot q - m)}{g}  - \frac{1}{(g - i q_0)} \\
& \times (i\gamma^k q_k + \gamma^4 q_0 - m) \bigg] \int^{\infty}_0 ds \bigg[ \frac{(2i q_0 g -g^2) }{(s + \vec{q}^2 -(q_0 + ig)^2)}   \\
&   \times \frac{\sum_1(-s)}{(s-m^2)}  - \frac{ i\gamma_4 g \sum_2(-s)}{s + \vec{q}^2 -(q_0 + ig)^2} \bigg] \bigg\}. 
\end{aligned}\end{equation}
The adiabatic limit  $g \rightarrow 0$ gives

\begin{equation}\begin{aligned}\label{ffaass}
&\langle 0 |\psi^{(r)} (x)  | q \rangle = \\
& \bigg[ 1 - \frac{1}{2} \int^{\infty}_0 ds \bigg(\frac{\sum_2(-s)}{s-m^2} - 2 m \frac{\sum_1(-s)}{(s-m^2)^2} \bigg) \bigg]\\
& \times \langle 0 |\psi^{(0)}(x)  | q \rangle,
\end{aligned}\end{equation}
where this term contains the wealth of information about the self-energy corrections of the electron amplitude. This operation relies on the fact that the free and asymptotic operators share the same dynamics and canonical commutation relations \cite{Yang}.

\begin{equation}\begin{aligned}\label{fazenda}
&Z_2^{-1/2} = \\
&1 - \frac{1}{2} \bigg[ \bar{\sum_2}(q^2)\bigg|_{q^2 \rightarrow - m^2} + 2 m \frac{ \partial \bar{\sum}_1 (q^2)}{\partial q^2}\bigg|_{q^2 \rightarrow - m^2} \bigg], 
\end{aligned}\end{equation}
where the counterterm $(Z_2 - 1) \equiv \delta_2$ at $e^2$ order is

\begin{equation}\begin{aligned}\label{elecpod}
\delta_2 &= \frac{e^2}{(4 \pi)^2 b^6} \bigg\{ 2 b^2 + 3 b^4 + 5 b^6 + (2 - 2 b^4 - 6 b^6 ) \ln{b} \ + \\
& \quad \frac{(1-2 b^2 - 3 b^4 - 6 b^6)}{\sqrt{(1-4 b^2)}}\bigg[ \ln{\bigg(\frac{b^4 - 2 b^2 + \sqrt{(1-4 b^2)}}{b^4 - 2 b^2 - \sqrt{(1 - 4 b^2)}}\bigg)} \\ 
& \quad - \ln{\bigg(\frac{1 + \sqrt{(1-4 b^2)}}{1 - \sqrt{(1 - 4 b^2)}}\bigg)} \bigg].
\end{aligned}\end{equation}
This correction term is equal to the value already found in \cite{Bufaloanomlous} if $\xi = 1$. Let us observe the $\delta_2$ and $\delta_m$ are solved exactly. The higher derivative term enforces the convergence of these counterterms. The $QED_4$ and $GQED_4$ counterterms are very different even if taking $m_P \rightarrow \infty$.


\subsection{Vertex correction $\Gamma_\mu$ }\label{trois}  

At this point, we proceed with the investigation of the $ \Gamma_\mu ( x'-x, x-x'')$ in \eqref{0nnn1}. The vertex correction arises two phenomena: anomalous magnetic moment and scale dependence of electric charge. These effects require the manipulation of the three-point correlation function whose properties explicitly preserved are the translation invariance and causality in the form

\vspace{1mm}

\begin{equation}\begin{aligned}\label{4433}
& \Gamma_\mu(x'-x, x-x'') = - \frac{e^2}{2} \gamma_\lambda \times \\
& \ \  tr[S^{(1)}(x'-x)\gamma_\mu S_{R}(x- x'') G_R(x''-x') \\ 
&  \  \quad + S_{A}(x'-x)\gamma_\mu S^{(1)}(x-x'') G_R(x''-x') \\ 
& \ \qquad + S_{A}(x'-x)\gamma_\mu S_R (x-x'') G^{(1)} (x''-x')] \gamma^\lambda.
\end{aligned}\end{equation}
Given the fact, we are working with a causality structure encoded by the retarded and advanced functions. This equation was first introduced by Schwinger in $(3+1)$ dimensions \cite{schwinger}. In the interaction picture, the splitting operation of the total Hamiltonian undermines a Fock representation to interacting operators. Otherwise, the $H$ vacuum would admit a unitary equivalence to the no-particle state of Fock space, which coincides with the $H_0$ vacuum. This statement means the Fock representation is unitarily equivalent to the interacting representation, and thus the interaction picture only describes trivial interactions. Nonetheless, Haag's theorem \cite{Haag} proves the Fock representation must be unitarily inequivalent to interacting operator representation because of $\mathcal{H}_{in,out} \cap \mathcal{H}_{int} = \O $. Therefore, it makes clear that the interacting operators have a non-Fock representation. Our strategy to overcome this foundation problem is to work out in the Heisenberg picture. First of all, we calculate the Fourier transform

 

\begin{equation}\begin{aligned}\label{cmd}
\Gamma_\mu (q,q') &= \int d^4 x' d^4 x'' \ e^{-iq(x' - x)}e^{-iq'(x-x'')} \\
& \qquad \qquad \ \qquad \quad \quad \times \Gamma_\mu(x'-x, x-x''). 
\end{aligned}\end{equation}
It is natural to expect many terms to vanish since the incoming and outgoing fermions obey the on-shell dispersion relation $q^2 = q'^2 = - m^2$. We shall separate this integral into the real and imaginary parts

\begin{equation}\label{cmddmc}
\Gamma_\mu (q,q') =  \Gamma^{(1)}_{\mu}(q,q') +  i \epsilon(q'-q) \Gamma^{(2)}_{\mu}(q,q'),   
\end{equation}
where the superficial degree of divergence is $-2$, as in Table \ref{table1}. This splitting solution leads to the real

\begin{widetext}

\hfill 

\begin{equation}\begin{aligned}\label{liu}
\Gamma^{(1)}_{\mu}(q,q') &= - \frac{e^2}{16 \pi^3} \int d^4 k  \bigg[ \mathcal{P} \frac{\delta^4 (k^2) - \delta^4 (k^2 + m_P^2) }{((q-k)^2 + m^2 )((q'-k)^2 + m^2 ) } + \mathcal{P} \frac{ m_P^2 \  \delta^4 ((q'-k)^2 + m^2) }{k^2(k^2 + m_P^2)((q-k)^2 + m^2)} \\
& \ \ \ + \mathcal{P} \frac{  m_P^2 \ \delta^4 ((q-k)^2 + m^2) }{k^2(k^2 + m_P^2)((q'-k)^2 + m^2) }\bigg]\gamma^\lambda(i\gamma \cdot (q-k)-m) \gamma_\mu(i\gamma \cdot (q'-k)-m)\gamma_\lambda, \\
\end{aligned}\end{equation}
and the imaginary part\footnote{The letter $\mathcal{P}$ indicates the principal value.} 
 
\begin{equation}\begin{aligned}\label{réspondre}
\Gamma^{(2)}_{\mu} (q,q') &= - \frac{e^2}{16 \pi^2} \int d^4 k  \delta^4((q - k)^2 + m^2)\delta^4((q' - k)^2 + m^2) \bigg(  \mathcal{P} \frac{1}{k^2} - \mathcal{P} \frac{1}{k^2 + m_P^2} \bigg) \\
& \ \ \ \times  [1 - \epsilon(q' - k)\epsilon(q- k)] \gamma^\lambda (i\gamma \cdot (q-k) - m)\gamma_\mu (i\gamma \cdot (q'-k)- m )\gamma_\lambda. \\
\end{aligned}\end{equation}
Let us start with the calculation of $\Gamma^{(2)}_{\mu}$. We get three different integral contributions using the product of $\gamma-$ matrices identities \cite{Bogoliubov}. The first one is the tensor integral which gives a rather complicated solution

\begin{equation}\begin{aligned}\label{nettoies}
& \mathcal{P} \int d^4 k \ \frac{ m_P^2 \ k_\mu k_\nu }{k^2(k^2 + m_P^2)}  \delta^4((q- k)^2 + m^2)\delta^4((q' - k)^2 + m^2) [1 - \epsilon(q' - k)\epsilon(q- k)]  =  \frac{\pi \Theta(-Q^2 - 4 m^2)}{Q^2 (1 + \frac{4m^2}{Q^2} )^{3/2} }    \frac{m_P^2}{Q^2}  \\
& \times \bigg\{ \bigg[ \bigg( 1 + \frac{4m^2}{Q^2} + \frac{ 3 m_P^2}{Q^2}\bigg) (  q'^\mu q'^\nu +  q^\mu q^\nu - g^{\mu\nu} \frac{Q^2}{2} ) - \bigg( 1 + \frac{4 m^2}{Q^2}  -  \frac{m_P^2}{Q^2}  \bigg) ( q'^\mu q^\nu + q^\mu q'^\nu + g^{\mu\nu} \frac{Q^2}{2} ) -     \\
&  g^{\mu\nu} 4 m^2 \bigg( 1 + \frac{4 m^2}{Q^2} + \frac{m_P^2}{Q^2} \bigg)  \bigg]  \frac{ \ln \bigg( 1 - \frac{Q^2  + 4 m^2}{m_P^2} \bigg)  }{(1+\frac{4 m^2}{Q^2}) } + \bigg[ \bigg( 1 + \frac{2 m^2}{Q^2} \bigg)(q^\mu q^\nu + q'^\mu q'^\nu - g^{\mu\nu} \frac{Q^2}{2}) - \frac{m^2}{Q^2} (q'^\mu q^\nu + \\
&  q^\mu q'^\nu  + g^{\mu\nu} \frac{Q^2}{2} ) -  \frac{m_P^2}{2} \bigg( 1 + \frac{4m^2}{Q^2} \bigg)   \bigg] \bigg\}, \\
\end{aligned}\end{equation}
\end{widetext}
where $Q = q'- q$ is the momentum transfer. Turning to the vector integral

\begin{equation}\begin{aligned}\label{eiporra3}
& \mathcal{P} \int \frac{  d^4 k \ k_\mu \ m_P^2 }{ k^2 (k^2 + m_P^2)} \delta^4((q- k)^2 + m^2)\delta^4((q' - k)^2 + m^2)  \\
&  \times  [1 - \epsilon(q' - k) \epsilon(q- k)] = \frac{\pi \Theta (- Q^2 - 4 m^2 )}{ Q^2 ( 1 + \frac{4 m^2}{Q^2})^{3/2} }  \frac{ m_P^2}{Q^2}   \\
&  \times \ln{\bigg( 1 - \frac{Q^2  + 4 m^2}{m_P^2} \bigg)}   ( q^\mu  + q'^\mu ). 
\end{aligned}\end{equation}
The infrared problem will appear when dealing with virtual photons. To handle the infrared catastrophe, we replace analytically $k^2 \rightarrow k^2 + \mu^2$. The scalar integral is 

\begin{equation}\begin{aligned}\label{xxaswq}
& \mathcal{P} \int d^4 k \ m_P^2 \frac{\delta^4((q- k)^2 + m^2)}{(k^2 + \mu^2)(k^2 + m_P^2) } \delta^4((q' - k)^2 + m^2) \\ 
& \times [1 - \epsilon(q' - k) \epsilon(q- k)] = \frac{ \pi \Theta(-Q^2 - 4 m^2)}{ -Q^2 \sqrt{1 + \frac{4m^2}{Q^2}}}  \\
& \times \bigg\{ \ln{ \bigg(1 - \frac{Q^2 + 4m^2}{\mu^2} \bigg)} -  \ln{ \bigg( 1 - \frac{Q^2 + 4 m^2}{ m_P^2 } \bigg)} \bigg\}. 
\end{aligned}\end{equation} 
Even though the vertex correction does not suffer from the ultraviolet divergence, the scalar integral reveals the photon introduces an infrared one.  In such a case, we add a fictitious mass $\mu$ for a well-behavior photon propagator at long distances. Then the lower limit of $k$ integration is divergence free
It is convenient to perform the limit of $\mu \rightarrow 0 $ to obtain physical results in the final solution. To understand the meaning of these integrals, we shall write the imaginary contribution \eqref{réspondre} following the process involved in the vertex correction

\begin{equation}\label{oreille}
\Gamma^{(2)}_{\mu}(q,q')  = \gamma_\mu F_1 (Q^2) \ + \ i \frac{(q + q')_\mu }{2 m} F_2(Q^2),
\end{equation}
where the $F_{1} (Q^2)$ is

\begin{widetext}
\begin{equation}\begin{aligned}\label{vertexf1}
F_1 (Q^2) &= - \frac{e^2}{8 \pi^2}\bigg\{ - \bigg( 1 + \frac{2 m^2}{Q^2} \bigg) \ln{\bigg( \frac{\mu^2 - (Q^2 + 4m^2)}{m_P^2 - (Q^2 + 4 m^2)}  \frac{m_P^2}{\mu^2} \bigg)} + \frac{3 m_P^2}{2 Q^2} \frac{\bigg[ 3 + \frac{4m^2}{Q^2} \bigg]}{ ( 1 + \frac{4m^2}{Q^2} ) } \\
& \ \quad  +  \frac{3 m_P^2}{Q^2} \bigg[ 1 + \frac{4m^2}{Q^2} - \frac{m_P^2}{Q^2} \bigg] \ln{\bigg( 1 - \frac{Q^2 + 4m^2}{m_P^2} \bigg) } \bigg\} \frac{\Theta(-Q^2 - 4m^2)}{ ( 1 + \frac{4m^2}{Q^2} )^{5/2} }, 
\end{aligned}\end{equation}
\end{widetext}
and $F_2(Q^2)$ is

\begin{equation}\begin{aligned}\label{vertexf2}
F_2(Q^2) &= \frac{e^2}{4 \pi^2} \frac{m_P^2}{Q^2} \frac{m^2}{Q^2} \frac{\Theta(-Q^2 - 4m^2)}{ ( 1 + \frac{4m^2}{Q^2} )^{3/2}} \\
& \quad \times \ln{\bigg( 1 - \frac{Q^2 + 4m^2}{m_P^2} \bigg) }.
\end{aligned}\end{equation}
These imaginary form factors correspond to the production and absorption of virtual particles. The $F_1$ represents the two virtual electron propagators, while the $F_2$ is the exchange of virtual photon by the two asymptotic electrons $| q \rangle$ and $| q' \rangle$. For our purposes, we shall relate \eqref{liu} with the same structure of \eqref{oreille}  to find the real form factor

\begin{equation}\label{lllabbeell}
\Gamma^{(1)}_{\mu}(q,q')  =  \gamma_\mu  \bar{F}_1(Q^2) + i \frac{ (q + q')_\mu }{2m} \bar{F}_{2}(Q^2).
\end{equation}
The well-known causality propriety of the singular functions in \eqref{4433} requires the $\Gamma_\mu$ lies on the retarded light-cone history, where this vector drops out if $x'_0 < x_0$ or $ x_0 > x''_0$. One can relate the real $ \Gamma_\mu^{(1)} $ and imaginary $\Gamma_\mu^{(2)}$ as the Kramers–Kronig relations \cite{Weinberg} since both of them are a Hilbert pair by means of analytic and translational invariance arguments

\begin{equation}\begin{aligned}\label{cacetttaaa}
\bar{F}_i (Q^2)&= \int^{\infty}_0 ds \frac{ F_i (-s)}{s + Q^2},    
\end{aligned}\end{equation}
where $i=\{ 1,2\} $ and $F_1(0) = 0$. The real component is evaluated in appendix \eqref{appa2}. We can now combine all radiative corrections calculated so far in \eqref{0nnn1}. By elementary manipulations, we rewrite each term to examine the results under the Heisenberg picture formalism. Firstly, we replace the first two terms of \eqref{0nnn1} by \eqref{label}. Secondly, the third term is equal to \eqref{cmddmc} with the aid of \eqref{xxaswq} and \eqref{lllabbeell}. Finally, the vacuum polarization was determined in \cite{Kallen1}. We thus end up with the operator current at second order

\begin{widetext}

\begin{equation}\begin{aligned}\label{préférer}
& \langle q | j^{(2)}_\mu (x) | q' \rangle = \langle q | j^{(0)}_\mu (x) | q' \rangle  \bigg[  -  \Bar{\Pi} (Q^2)  +  \Bar{\Pi} (0)  -  i   \epsilon (Q) \Pi (Q^2) -  \bar{\sum_{2}} (-m^2)  -  2 m \bar{\sum_{1}}' (-m^2) \\
&  + \bar{F}_{1  } (Q^2) +  i   \epsilon(Q) F_{1  } (Q^2) \ \bigg]  - \frac{e}{2m} (q+q')_\mu  \bigg[ \ \bar{F}_{ 2 } (Q^2) + i   \epsilon(Q) F_{2  } (Q^2) \ \bigg] \langle  q | : \bar{\psi}^{(0)} (x) \psi^{(0)} (x) : | q' \rangle,
\end{aligned}\end{equation}

\end{widetext}
where the polarization is encoded by $\Pi$. Following the Lagrangian \eqref{asdfgfds}, the polarization effect is left unchanged because the minimal coupling and fermion particles give rise to the same result except for the $QED_4$. To get a qualitative interpretation, we emphasize the correction is compound by two terms. The former means a multiplicative factor proportional to $\langle q|j^{(0)}_\mu (x)| q' \rangle$. The latter term, responsible for the anomalous magnetic moment, will be analyzed in the next section\footnote{$\bar{\sum_{1}}' (-m^2) = \partial \bar{\sum}_1 (q^2) / \partial q^2|_{q^2 \rightarrow - m^2}$.}.

Since the bare and dressed vacuum share the same domain in the interaction picture, the existence of Fock representation is spoiled for $H_{int}$  \cite{Seidewitz}. On the other hand, the Heisenberg picture finds a way to bypass such a problem because of defining the dynamical of the interacting and free operator in the same Hilbert space. With this in mind, interacting and asymptotic operators continue having the same Fock representation \cite{Yang,Bogoljubov2}. The free, "in", and "out" operators satisfy the same commutation relations \cite{Yang}, we can therefore implement a linear map connecting each one at different time by a unitary transformation.

\section{The magnetic moment of electron}\label{fin}

The satisfactory formulation of $QED_4$ occurred when Schwinger deduced the deviations of electron magnetic moment from the Dirac equation \cite{schwinger}. We can now achieve higher precision using a numerical calculation for the higher-order loops \cite{amms}. In recent years, the experimental investigation has showed the anomalous magnetic moment of the electron is the most accurately value in physics $ a_{exp} = 1,15965218073 \times 10^{-3} \pm 2,8 \times 10^{-13} $ \cite{evamm}. A look in the last term of \eqref{préférer} gives a direct interpretation of how this process arises. We start by rewriting this expression with the help of Gordon's identity

\begin{equation}\begin{aligned}\label{mettent}
& \frac{e}{2m} (q+q')_\mu \langle  q |:  \bar{\psi}^{(0)} (x) \psi^{(0)} (x): | q' \rangle =  \\
& \langle  q | j_\mu^{(0)} (x) | q' \rangle + \frac{i e}{2m} Q^\nu \langle q|: \bar{\psi}^{(0)}(x) \sigma_{\mu\nu} \psi^{(0)}(x):|q' \rangle, 
\end{aligned}\end{equation}
where $\sigma_{\mu\nu} = \frac{i}{2} (\gamma_\mu \gamma_\nu - \gamma_\nu \gamma_\mu)$. According with the previous calculation, the electron obtains a shift in the magnetic moment

\begin{equation} 
\frac{e}{2m} \langle  q | :   \bar{\psi}^{(0)} (x) \sigma_{\mu\nu} \psi^{(0)}(x) : | q' \rangle \bigg[ \bar{F}_{2} (Q^2)  + i  \epsilon(Q) F_{2} (Q^2) \bigg].
\end{equation}
Let us retain the first order of the total magnetic moment for a one-electron state. Taking $ Q \rightarrow 0 $, we arrive at 

\begin{equation} 
\frac{e}{2m} \langle q|: \bar{\psi}^{(0)}(x) \sigma_{\mu\nu}\psi^{(0)} (x) : | q \rangle \bigg[ 1 +  \bar{F}_{2}(0)  \bigg],
\end{equation}
where the $\bar{F}_{2}(0)$ is a convergent quantity

\begin{equation}\begin{aligned}\label{countmag}
&\bar{F}_{2} (0) = \int^{\infty}_0 \frac{d s}{s}  F_2 (-s) =   \\
&   \frac{e^2}{4 \pi^2 b^4}  \bigg\{   b^2 + (1-2 b^2) \ln{b}  - \frac{b^2}{2} - \bigg( \frac{2 b^4 + 1 - 4 b^2 }{ 2 b^4 \sqrt{1- 4 b^2}} \bigg) \times \\
& \bigg[ \ln{\bigg( \frac{ \sqrt{1- 4 b^2 } + 1}{\sqrt{1-4 b^2} - 1} \bigg) } + \ln{\bigg( \frac{\sqrt{1-4 b^2} - 1 + 2 b^2 }{ \sqrt{1-4 b^2} + 1 - 2 b^2} \bigg) } \bigg] \bigg\}.
\end{aligned}\end{equation}
This result emerges when the $QED_4$ fails to describe the physical phenomena exploited in \cite{deuxdeux,unun}. We are mainly interested in the explicit change of the finite anomalous magnetic moment arising without any standard regularization procedure. If we interpret this solution as the measurement error from the anomalous magnetic moment, we can settle down a restriction for the Podolsky parameter $ m_P \geq 3,7595 \times 10^{10} \ eV $. From the above result, we find the adequacy between this value with the Interaction picture in \cite{Bufaloanomlous}. However, we may not assure this validity for higher order terms in the perturbative method or non-perturbative one.


\section{Conclusions and Outlook}\label{doendo}

Our theory shed new light on exploring the spinor-photon coupling. We studied the efficiency of this higher derivative theory to prevent ultraviolet problems in quantum electrodynamics. The stability, unitarity, and same symmetries from the $QED_4$ are the features that give further support for our framework. Before focusing on direct application, we guaranteed the $GQED_4$ was not part of the effective field theory class but rather a unitary extension of the $QED_4$ \cite{Pimente}. Among all possible combinations, the $GQED_4$ is far from being redundant as pointed by \cite{Cuzinattogauge}. With a view towards a physical description, our approach technically demanded a gauge fixing update. If this subject has not been done correctly, it would subsequently reproduce questionable results as ghost states. To restore the canonical quantization and assure contact with bounded operators, we used the non-mixing gauge in \cite{nomixing} instead of the Lorentz gauge. Then we could write down the spectral support for higher energy regions. As argued in \cite{bufalopath}, the $QED_4$ cannot be recovered by taking $m_P \rightarrow \infty $ in \eqref{countmass}, \eqref{elecpod}, \eqref{préférer} and \eqref{countmag} because the gauge operator could not split into two parts contributions as in \eqref{cachorro}. Consequently, this operation led us to run out the spectral condition and consequently the Lorentz covariance.

Applying these methods developed so far, we computed the $\Gamma^\mu$ and $\sum$ corrections at $e^2$ order. We confirmed the $GQED_4$ was favorable to a divergenceless scenario if we adopt a geometric deformation for the covariant photon propagator \eqref{sssa}. The functions $\Gamma^\mu$ and $\sum$ converge to a finite value at $1-$loop order despite the ultraviolet divergences of the polarization $\Pi^{\mu\nu}$. We should renormalize the electric charge because the Podolsky parameter only concerns the photon sector rather than the fermion one. As expected, the gauge invariant cancellation in the ultraviolet region left off the infrared divergences. Even though we may tempt to guess that $m_P$ plays an infrared regulator as the electron mass, the infrared excitations, roughly speaking, are unknown or agnostic for higher derivative terms. The enlargement of the parameter space due to $GQED_4$ cannot govern the behavior of the low-energy properties.

At the end of this paper, we opened the possibility to test the $GQED_4$ corrections by investigating the next-to-leading order contribution to the anomalous magnetic moment. Significant efforts have been made to cure the discrepancy between experimental and theoretical predictions. For example, the CERN \cite{exp1exp} and BNL E821 \cite{exp2exp} substantially improve the measurement of muon magnetic anomaly. Using the data in \cite{evamm}, we could constrain the parameter $ m_P \geq 37.595 \ GeV $. As we know, the $QED_4$ leads the contribution concerning electroweak and quantum chromodynamics. Our choice is satisfactory since the high light photon from $GQED_4$ cannot be ignored when electroweak (Z-boson) \cite{Czarnecki1995sz}, hadronic vacuum polarization \cite{Gerardin2020gpp}, and hadronic light-by-light \cite{liglig} are included. In addition, the $QED_4$ contribution up to tenth-order remains insignificant to discrepancy between Standard Model and experiments \cite{tenthQED}. The $GQED_4$ can provide a significant advantage to create a Laboratory to detect the new structures emerging from a high-energy photon. The Podolsky parameter, playing the central role to get rid of the ultraviolet divergences, extends the validity of $QED_4$ to $m^2 \leq p^2 \leq m_P^2$.  We also test Podolsky's theory perturbatively by spectroscopy of Hydrogen atom \cite{CUZINATTO} and Bhabha scattering \cite{Bufalo2014jra}. With this in mind, these cutoffs in different energy regimes provide different laboratories to detect the new extensions for quantum electrodynamics.

We have discussed the main problems raised by the interaction picture to quantum fluctuations without appealing to any axiomatic explanation. What have we learned from the Heisenberg picture? The analysis put forward showed the perturbative expansion contains the well-defined self-adjoint operator at different times \cite{chatooo}. In such a representation, we could define a linear map for creation and annihilation operators at all times. Such assumption allowed us to express the operators into a series expansion in the power of gauge coupling. Again, we showed the radiative correction at $e^2$ order in \eqref{elecpod} yields an accurate prediction when compared with the interaction picture from \cite{Bufaloanomlous}. We stress that the one-to-one correspondence (unitary equivalence) between the interaction and the Heisenberg picture is absent in the perturbative and non-perturbative methods. Our analyzes suggest the study of the non-perturbative approach, where a discrepancy should appear. It will be published elsewhere.

We hope that the framework proposed at $T=0$ can provide new insights into finite temperature. The corrections to Stefan-Boltzmann's law paved the road to evaluate possible observable quantities. In particular, one should expect to determine the Podolsky parameter from the cosmic microwave background data due to the polarization \cite{cmbcmb}.  Beyond that, the $GQED_4$ will be helpful to investigate the curved space-time \cite{Zayats} and self-interaction in an accelerated frame \cite{Gratus}. We will postpone these discussions for future work.

The $GQED_4$ is a natural (unitary and stable) treatment to control the ultraviolet regime. Due to work in \cite{15koa}, the functions $\sum$ and $\Gamma$ are ultraviolet finite at-any loop order and $\Pi$ for two loops and so on. Since $GQED_4$ is a super-renormalizable theory, the superficial degree of divergence decrease by $2$ orders from $n$ to $n+1$ loops \cite{15koa}. Therefore, excepting the $\Pi$ is logarithmic divergent at $1-$loop, the $GQED_4$ is free from ultraviolet divergence.

At present, we cannot offer a systematic procedure to cure the ultraviolet divergence of vacuum polarization at $e^2$ order based on the $GQED_4$. We end by noting that other clean recipes can extend our formalism. From the work in \cite{vtnc1}, it is possible to introduce an effective mass to obtain an asymptotically free theory in the high energy regime. It would be interesting to incorporate new techniques to create a safe asymptotic regime in vacuum polarization to $GQED_4$. The arguments discussed in \cite{vtnc2} looked in favor of the non-local extension to restore the ultraviolet finite by dominating the propagator behavior. We also found the inclusion of non-local higher derivative term turns finite the Einstein–Hilbert action at quantum level \cite{vtnc3}.


\vspace{2mm}

\noindent
{\em Acknowledgements.}
David Montenegro thanks to CAPES for full support. 


\appendix

\section{Electron transition amplitude}\label{appa}

Here, we formally solve the operator \eqref{firstese} in detail
 
\begin{equation}\label{0234}
 \Phi(x) = \int d^4x'  \sum(x-x') \psi^{(0)} (x').   
\end{equation}
By using \eqref{RRpsi} at $e^2$ order in \eqref{firstese}, we write down the operator in terms of singular functions

\begin{equation}\begin{aligned}\label{zorra}
\sum(x-x') &= - \frac{e^2}{2} \gamma_\lambda [S^{(1)}(x-x') G_A (x'-x)  \\
&  \qquad \quad \ +  S_{R}(x-x') G^{(1)} (x'-x)  ] \gamma^\lambda .
\end{aligned}\end{equation} 
Now we denote the Fourier transform of $\sum(x-x')$ as 

\begin{equation}
\sum (x-x') = \int \frac{d^4 q}{(2 \pi)^4} e^{iq(x-x')} \sum (q),   
\end{equation}
where $\sum (q)$ is a retarded function, which vanishes outside the light cone, 
   
\begin{equation}\begin{aligned}\label{realone}
& \sum (q) = \\
& \frac{- e^2}{2} \int \frac{d^4 k}{(2\pi)^3}   \gamma^\lambda (i\gamma \cdot (q + k) -m) \gamma_\lambda   \bigg[ \delta^4( (q + k)^2 + m^2)  \\
& \times \bigg( \mathcal{P}\frac{1}{k^2} - \mathcal{P} \frac{1}{k^2 + m_P^2} -  i \pi \epsilon(k) (  \delta^4 (k^2) - \delta^4 (k^2 + m_P^2) ) \bigg)  \\ 
&  + \bigg( \mathcal{P} \frac{1}{(q + k)^2 + m^2 } + i \pi \epsilon(q+k) \delta^4((q + k)^2 + m^2) \bigg) \\ 
&  \times (\delta^4 (k^2) - \delta^4 (k^2 + m_P^2) ) \bigg]. 
\end{aligned}\end{equation}
This integral has the superficial degree of divergence equal to $-1$, as in Table \ref{table1}. If we calculate higher orders in the electron-self energy (more loops are added), the superficial degree of divergence decrease because the $GQED_4$ is a super-renormalizable theory. Now we are motivated to separate the integral and take the imaginary part.

\begin{equation}\begin{aligned}\label{demandé}
& Im \sum(q) = \\ 
&\frac{- e^2}{ 2 } \int \frac{d^4 k}{(2\pi)^2}     \gamma^\lambda  ( i\gamma \cdot (q + k)  - m)\gamma_\lambda \delta^4 ((q + k)^2 + m^2)      \\ 
&  \times    (\delta^4 ( k^2) - \delta^4 ( k^2 + m_P^2))   ( \epsilon(  q+ k  ) - \epsilon( k ) ). 
\end{aligned}\end{equation}
We begin by assuming the solution has a simple form

\begin{equation}\label{caralho}
Im \sum(q)  = \epsilon (q) \bigg[ \sum_1 (q^2)  + (i \gamma \cdot q + m) \sum_2 (q^2) \bigg].
\end{equation}
We express the off-shell photon and fermion particles in terms of the imaginary contribution from $\sum_1$ and $\sum_2$, respectively. To calculate each component separately, we can invoke the identities of $\gamma-$matrices. If we multiply $\gamma_\mu$ and take the trace, we get $  tr [\gamma_\mu  Im \sum(q) ] =  2 i q_\mu \epsilon (q) \sum_2 (q^2) $. Then

\begin{equation}\begin{aligned}\label{lam}
& \epsilon (q) \sum_2 (q^2) = \\
& \frac{e^2}{16 \pi^2}  \int  d^4 k \bigg(1- \frac{m^2}{q^2} \bigg) \delta^4(q^2 + 2qk + k^2 + m^2) \\
& \times ( \delta^4(k^2) - \delta^4(k^2 + m_P^2) ) (\epsilon(q+k) - \epsilon(k)), 
\end{aligned}\end{equation}
which we can easily solve

\begin{equation}\begin{aligned}\label{f2o}
&\sum_2 (q^2) = \\ 
& \frac{e^2}{16 \pi} \bigg\{ \bigg( 1 -  \frac{m^4}{q^4}  \bigg) \Theta(-q^2 -m^2) - \bigg( 1-\frac{m^2 - m_P^2}{q^2} \bigg) \\ 
& \times \sqrt{  \bigg( 1+\frac{m^2 - m_P^2}{q^2} \bigg)^2 +\frac{4m_P^2}{q^2} } \Theta(-q^2 - (m + m_P)^2) \bigg\}.
\end{aligned}\end{equation}
In a quite analogous way, taking the trace of \eqref{caralho} $ tr [ Im \sum (q) ] =  4 \epsilon (q) ( \sum_1 (q^2) + m \sum_2 (q^2) ) $. Then , we obtain

\begin{equation}\begin{aligned}
&\epsilon (q) \bigg[ \sum_1 (q^2) + m \sum_2 (q^2) \bigg] = \\
& \frac{m e^2}{4 \pi^2} \int d^4 k  \delta^4((q+k)^2 + m^2) (\delta^4(k^2) -  \delta^4(k^2 + m_P^2) ) \\ 
& \times  (\epsilon(q+k) - \epsilon(k)), 
\end{aligned}\end{equation}
which results

\begin{equation}\begin{aligned}\label{gratuits}
&\sum_1 (q^2) = \\
&\frac{m e^2}{16 \pi^2} \bigg\{ \bigg( 1 + \frac{m^2}{q^2} \bigg) \bigg( 3 + \frac{m^2}{q^2} \bigg) \Theta(-q^2 -m^2) - \\
&  \bigg( 3 + \frac{m^2 - m_P^2}{q^2} \bigg)\sqrt{\bigg(1+\frac{m^2 - m_P^2}{q^2}\bigg)^2 +\frac{4m_P^2}{q^2}}
 \\
& \  \times \Theta(-q^2 - (m + m_P)^2) \bigg\}.
\end{aligned}\end{equation}
It is possible to construct the real component in \eqref{realone}, we introduce the Kramers–Kronig relations \cite{Weinberg}. To investigate this aspect, the linear operator satisfy the properties of causality by vanishing in the advanced light cone $(x_0 < x'_0)$ and analyticity

\begin{equation}\label{lajel}
\bar{\sum_{i}} (q^2) = \mathcal{P} \int^{\infty}_{0} ds \frac{\sum_{i} (-s)}{s+q^2},      \\ \\ 
\end{equation}
where $i = \{ 1,2 \} $. Then we can rewrite $\sum (q)$ as

\begin{equation}\begin{aligned}\label{sdwer}
\sum (q) =  & \bar{\sum_{1}}(q^2)  +  i \epsilon(q) \sum_{1}(q^2) +   \\
&  (i \gamma \cdot  q + m) \bigg[  \bar{\sum_{2}} (q^2) + i \epsilon(q) \sum_{2}(q^2) \bigg]. \\ \\
\end{aligned}\end{equation}
This function encodes the photon and electron propagator in the Heisenberg picture. The terms $\bar{\sum_{1}}(q^2)$ and $\sum_{1}(q^2)$ concern the virtual Podolsky photon and $\bar{\sum_{2}}(q^2)$ and $\sum_{2}(q^2)$ are the virtual fermion. Substituting \eqref{lam}, \eqref{gratuits} and \eqref{lajel} in \eqref{sdwer}, we obtain

\begin{widetext}
\begin{equation}\begin{aligned}\label{0htrans11}
\sum (q) &= \frac{m e^2}{16 \pi^2 }  \bigg\{ \bigg[  \frac{m^2}{q^2} + \bigg( 3 + \frac{4 m^2}{q^2} + \frac{m^4}{q^4}\bigg) \ln{\bigg( 1 + \frac{q^2}{m^2} \bigg)} + 3 \frac{4m m_P}{q^2}\bigg( \frac{m + m_P }{m - m_P} \bigg)^2 \ln{\bigg(\frac{2m m_P}{m^2 + m_P^2}\bigg)} + \bigg( 3 + \frac{m^2 - m_P^2}{q^2}  \bigg)  \\
& \quad \times \sqrt{\frac{q^2 + (m+m_P)^2}{q^2 + (m-m_P)^2}}  \ln{\bigg( \frac{1- \sqrt{\frac{q^2 + (m-m_P)^2}{q^2 + (m+m_P)^2}}}{1+ \sqrt{\frac{q^2 + (m-m_P)^2}{q^2 + (m+m_P)^2}}} \bigg)} \bigg]  + i \pi   \epsilon(q) \bigg[ \bigg( 1 + \frac{m^2}{q^2} \bigg)\bigg( 3 + \frac{m^2}{q^2} \bigg) \Theta(-q^2 -m^2)  \\
& \quad   - \bigg( 3 + \frac{m^2 - m_P^2}{q^2} \bigg)\sqrt{\bigg(1+\frac{m^2 - m_P^2}{q^2}\bigg)^2 +\frac{4m_P^2}{q^2}} \ \Theta(-q^2 - (m + m_P)^2) \bigg] \bigg\} + (i \gamma q + m) \frac{ e^2}{16 \pi^2 } \bigg\{ \bigg[  - \frac{m^2}{q^2} \\
& \quad + \bigg(1  - \frac{m^4}{q^4} \bigg) \ln{\bigg( 1 + \frac{q^2}{m^2} \bigg) }    -  \frac{4 m m_P}{q^2} \bigg( \frac{m+m_P}{m-m_P} \bigg)^2 \ln{\bigg(\frac{2m m_P}{m^2 + m_P^2}\bigg)} - \bigg( 1 - \frac{m^2 - m_P^2}{q^2}  \bigg)   \sqrt{\frac{q^2 + (m+m_P)^2}{q^2 + (m-m_P)^2}} \times \\
& \quad  \ln{\bigg( \frac{1- \sqrt{\frac{q^2 + (m-m_P)^2}{q^2 + (m+m_P)^2}}}{1+ \sqrt{\frac{q^2 + (m-m_P)^2}{q^2 + (m+m_P)^2}}} \bigg)} \bigg] + i \pi \epsilon(q) \bigg[ \bigg( 1 -  \frac{m^4}{q^4}  \bigg) \Theta(-q^2 -m^2)  - \bigg( 1-\frac{m^2 - m_P^2}{q^2} \bigg) \sqrt{  \bigg( 1+\frac{m^2 - m_P^2}{q^2} \bigg)^2 +\frac{4m_P^2}{q^2} }   \\
& \quad \times \Theta(-q^2 - (m + m_P)^2) \bigg] \bigg\}. 
\end{aligned}\end{equation}
\end{widetext}

\section{Calculus of $\bar{F}_1 (Q^2)$ and $\bar{F}_2 (Q^2)$ }\label{appa2}

\vspace{3mm}

To evaluate the real components of vertex function $\bar{F}_1 (Q^2)$ and $\bar{F}_2 (Q^2)$, we will follow the Kramers–Kronig relations in \eqref{cacetttaaa}. The derivation begins with the imaginary component in \eqref{vertexf1}. The real part of the electric charge scale dependence

\begin{equation}\begin{aligned}\label{b1p}
& \bar{F}_1 (Q^2) = \frac{(e m_P)^2}{8 \pi^2} \int^{\infty}_{4m^2}  \frac{ds}{(s + Q^2)} \frac{1}{(1- \frac{4m^2}{s})^{1/2}} \times  \\ 
& \bigg[ \frac{( s - 2 m^2 )}{(s - 4 m^2)^2}   \ln{\bigg( \frac{s - 4m^2 + \mu^2 }{s - 4 m^2 + m_P^2}  \frac{m_P^2}{\mu^2} \bigg)}  + \frac{3s}{2} \frac{( 3 s - 4m^2 )}{ ( s - 4m^2 )^3 }    \\
& + \frac{3 ( s - 4m^2 + m_P^2)}{(s - 4m^2)^2} \ln{\bigg( 1 + \frac{ s  - 4m^2}{m_P^2} \bigg) } \bigg],
\end{aligned}\end{equation}
By dimensional analysis, it is easy to see the superficial degree of divergence is $-2$, as in Table \eqref{table1}, and therefore, we obtain finite integral due to Podolsky contribution. Carrying out the same steps as in the previous calculation, the rules for the real component in \eqref{vertexf2} is

\begin{equation}\begin{aligned}\label{b2p}
\bar{F}_2 (Q^2)=& \frac{ ( e m_P )^2}{ 4 \pi^2}  \int^{\infty}_{4m^2}  \frac{ds}{(s + Q^2)} \frac{m^2}{\sqrt{s}} \frac{1}{ ( s - 4m^2)^{3/2}} \\
& \times \ln{\bigg( 1 + \frac{s - 4m^2}{m_P^2} \bigg) }. \\
\end{aligned}\end{equation}
The integrals in Eqs \eqref{b1p} and \eqref{b2p} are calculated by elementary methods where we assume $ 1 + 4 m^2 / Q^2 > 0$.

\section{The superficial degree of divergence of the $GQED_4$}\label{apdc}

\vspace{2mm}

We have known the mass dimensionality of the interaction coefficient $(\Delta)$ is responsible for determining the divergence fate in a scattering amplitude \cite{Weinberg}. The message from this dimensionality is that we have three possibles theories: non-renormalizable $( \Delta < 0 )$, renormalizable $(\Delta = 0)$, and super-renormalizable $( \Delta > 0 )$. We need to keep in mind that the superficial degree of divergence $(D)$ corresponds to the upper bound order of integral ultraviolet divergence. From these descriptions found in quantum books, we would suppose the $GQED_4$ is a renormalizable theory because the gauge coupling of $ j^\mu A_\mu $ has a mass dimension $0$ (marginal). Thus, contrary to the statements above, our theory is not renormalizable but rather super-renormalizable. Then the standard classification is only consistent with Lagrangians described by first order derivative at the kinematic level. To higher derivative theories, we use the super-renormalizable definition found in \cite{Tomboulis}.


\begin{table}[h!]
\begin{center}
\begin{tabular}{ |c|c|c|c| } 
\hline
 & $QED_4$ & $GQED_4$ &  \\
\hline
\hline Vacuum Polarization & 2 & 2 & $D$ \\ 
Vertex Correction & 0 & -2 & $D$ \\ 
Electron self-energy & 1 & -1 & $D$ \\ 
\hline
\end{tabular}
\caption{\label{tabel1} The superficial degree of divergence for one-particle irreducible at $e^2$ order.}
\label{table1}
\end{center}
\end{table}
The $D$ of $\Pi$, $\sum$, and $\Gamma$ decreases $-2$ order from $n$ to $n + 1$ loops \cite{15koa}. Therefore, the $GQED_4$ one-particle irreducible becomes more convergent for higher orders loops.





\end{document}